\documentclass{pasj00}

\begin{document}
\SetRunningHead{T. Kato et al.}{Grazing Eclipsing Dwarf Nova CW Monocerotis}

\Received{}
\Accepted{}

\title{Grazing Eclipsing Dwarf Nova CW Monocerotis: Dwarf Nova-Type Outburst
       in a Possible Intermediate Polar?}

\author{Taichi \textsc{Kato} and Makoto \textsc{Uemura}}
\affil{Department of Astronomy, Kyoto University,
       Sakyo-ku, Kyoto 606-8502}
\email{tkato@kusastro.kyoto-u.ac.jp, uemura@kusastro.kyoto-u.ac.jp}

\author{Seiichiro \textsc{Kiyota}}
\affil{Variable Star Observers League in Japan (VSOLJ),
       1-401-810 Azuma, Tsukuba 305-0031}
\email{skiyota@nias.affrc.go.jp}

\author{Kenji \textsc{Tanabe}, Mitsuo \textsc{Koizumi}, Mayumi \textsc{Kida},
       Yuichi \textsc{Nishi}, Sawa \textsc{Tanaka}, Rie \textsc{Ueoka},
       Hideki \textsc{Yasui}}
\affil{Department of Biosphere-Geosphere Systems,
       Faculty of Informatics, Okayama University of Science, \\ Ridaicho 1-1,
       Okayama 700-0005}
\email{tanabe@big.ous.ac.jp}

\author{Tonny \textsc{Vanmunster}}
\affil{Center for Backyard Astrophysics (Belgium), Walhostraat 1A, B-3401
       Landen, Belgium}
\email{Tonny.Vanmunster@cbabelgium.com}

\author{Daisaku \textsc{Nogami}}
\affil{Hida Observatory, Kyoto University, Kamitakara, Gifu 506-1314}
\email{nogami@kwasan.kyoto-u.ac.jp}

\email{\rm{and}}

\author{Hitoshi \textsc{Yamaoka}}
\affil{Faculty of Science, Kyushu University, Fukuoka 810-8560}
\email{yamaoka@rc.kyushu-u.ac.jp}

\KeyWords{
          accretion, accretion disks
          --- stars: binaries: eclipsing
          --- stars: dwarf novae
          --- stars: individual (CW Monocerotis)
          --- stars: novae, cataclysmic variables
          --- stars: oscillations
}

\maketitle

\begin{abstract}
   We observed the 2002 October--November outburst of the dwarf nova CW Mon.
The outburst showed a clear signature of a premaximum halt, and a more
rapid decline after reaching the outburst maximum.  On two separate
occasions, during the premaximum stage and near the outburst maximum,
shallow eclipses were recorded. This finding confirms the previously
suggested possibility of the grazing eclipsing nature of this system.
The separate occurrence of
the eclipses and the premaximum halt can be understood as
a result of a combination of two-step ignition of an outburst and the
inside-out propagation of the heating wave.  We detected a coherent
short-period (0.02549 d) signal on two subsequent nights around the optical
maximum.  This signal was likely present during the maximum phase
of the 2000 January outburst.  We interpret this signal as a signature of
the intermediate polar (IP) type pulses.  The rather strange outburst
properties, strong and hard X-ray emission, and the low luminosity of
the outburst maximum might be understood as consequences of the supposed
IP nature.  The ratio between the suggested spin period and the orbital
period, however, is rather unusual for a system having an orbital period
of $\sim$0.176 d.
\end{abstract}

\section{Introduction}

   Dwarf novae are a class of cataclysmic variables \citep{war95book}.
Dwarf novae show outbursts, which are believed to be a result of the
disk-instabilities in the accretion disk \citep{osa96review}.
The eclipses in some high-inclination dwarf novae provide a unique tool
in studying the time-evolution of the geometry and physical properties
of the accretion disk (e.g. \cite{EclipseMapping}; \cite{woo89ippeg};
\cite{wol93ippeg}; \cite{web99ippeg}; \cite{ioa99htcas}).  The appearance
of eclipses in grazing eclipsing systems can provide a powerful tool
in studying the radius change in an outbursting disk
(e.g. \cite{sma84ugemdiskradius}), which is essential for distinguishing
the outburst mechanisms \citep{ich92diskradius}.

   CW Mon was originally discovered by \citet{ahn44cwmon}, whose observation
(also shown in \cite{GlasbyDNbook}) demonstrated the dwarf nova-type
variability.  The object was photographically studied by \citet{wac68cwmon}.
The exact identification was independently studied by several authors
(\cite{vog82atlas}; \cite{lop85CVastrometry}; \cite{SkiffIDibvs4676};
\cite{KinnunenIDibvs4863}).  \citet{bat97cwmon} reported that mean outburst
cycle length is $\sim$160 d, although many outbursts must have been missed
due to unavoidable seasonal gaps.  \citet{stu97cwmon} further presented
an analysis of the outburst statistics, and concluded that CW Mon has
wide and narrow outbursts, the former being $\sim$0.5 mag brighter than
the latter.  \citet{stu97cwmon} proposed a mean cycle length of $\sim$150 d
with a statistical assumption of missed outbursts.

   \citet{szk86cwmonxleoippegafcamIR} obtained time-resolved infrared
photometry of this object, and detected ellipsoidal variations attributable
to a binary motion with a period of $P_{\rm orb}$ = 4.23$\pm$0.01 hr.
\citet{szk86cwmonxleoippegafcamIR} further noted the possible presence of
a shallow fading which could be attributed to a grazing eclipse of the
accretion disk.  \citet{szk86cwmonxleoippegafcamIR} deduced an inclination
angle of $i$ = 65$^{\circ}$ from the infrared light curve and the profile
analysis of the Balmer emission lines.  \citet{how88faintCV1} obtained
an optical time-series photometry and detected a possible dip, which
may be associated with an eclipse.  However, the identification of the
nature of the dip remained unclear because of the lack of the clearly
recurring nature of this phenomenon \citep{how88faintCV1}.

   From high-speed CCD photometry, \citet{sto87highspeedCCD} reported the
possible presence of an eclipse lasting for $\sim$1900 s.  Although
there is a report \citep{sto87cwmonbaas} that these eclipses {\it sometimes}
occur, especially during the state following an outburst, neither detailed
nor systematic observations have yet been published.

   We conducted time-resolved CCD observing campaign through the VSNET
Collaboration,\footnote{
  $\langle$http://www.kusastro.kyoto-u.ac.jp/vsnet/$\rangle$
} upon the alerts of outburst detection in 2002 October--November by
D. Taylor and M. Simonsen, vsnet-outburst 4679, 4683\footnote{
  $\langle$http://www.kusastro.kyoto-u.ac.jp/vsnet/Mail/\\outburst4000/msg00679.html$\rangle$, \\
  $\langle$http://www.kusastro.kyoto-u.ac.jp/vsnet/Mail/\\outburst4000/msg00683.html$\rangle$
}).  We also obtained some data during the 2000 January outburst, which was
detected by H. McGee, vsnet-alert 3952.\footnote{
  $\langle$http://www.kusastro.kyoto-u.ac.jp/vsnet/Mail/\\alert3000/msg00952.html$\rangle$
}

\begin{figure*}
  \begin{center}
    \FigureFile(160mm,80mm){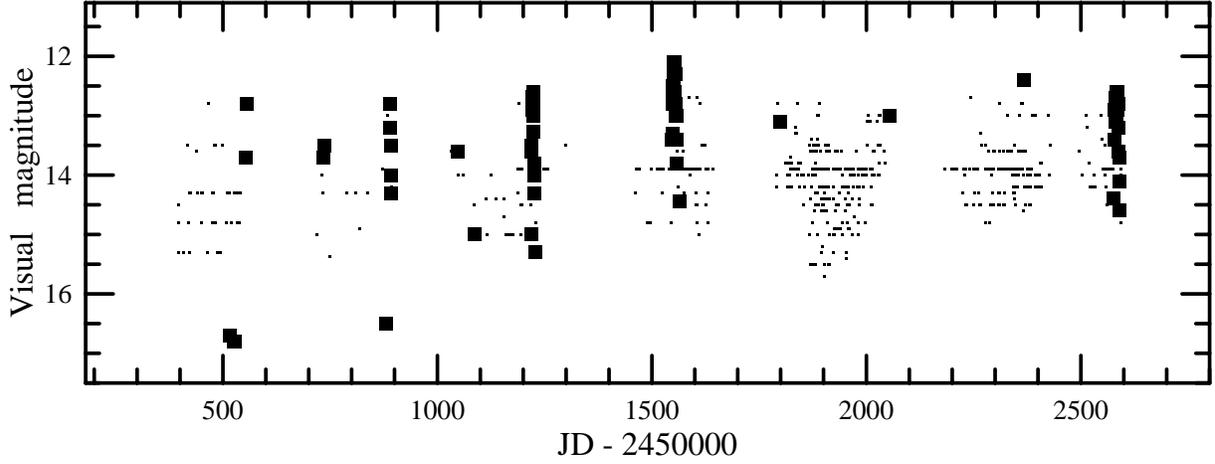}
  \end{center}
  \caption{Long-term visual light curve of CW Mon constructed from the
  observations reported to VSNET.  Large and small dot represent positive
  and negative (upper limit) observations, respectively.  Outbursts rather
  rarely occur once in 100--200 d.
  }
  \label{fig:long}
\end{figure*}

\begin{figure}
  \begin{center}
    \FigureFile(88mm,60mm){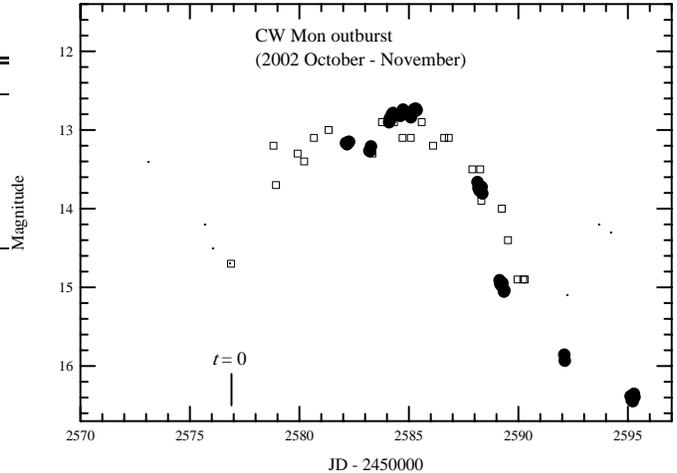}
  \end{center}
  \caption{Overall light curve of the 2002 October--November outburst.
  The large filled circles and open squares represent CCD (this observation)
  and visual (reported to VSNET) observations, respectively.
  The small dots represent upper limit visual observations.  The magnitude
  scale was adjusted to $V$.  CCD observations represent averaged magnitude
  in 0.05-d bins.  A 0.3 magnitude was added to the visual
  observations in order to correct the systematic difference.}
  \label{fig:out}
\end{figure}

\begin{figure}
  \begin{center}
    \FigureFile(88mm,60mm){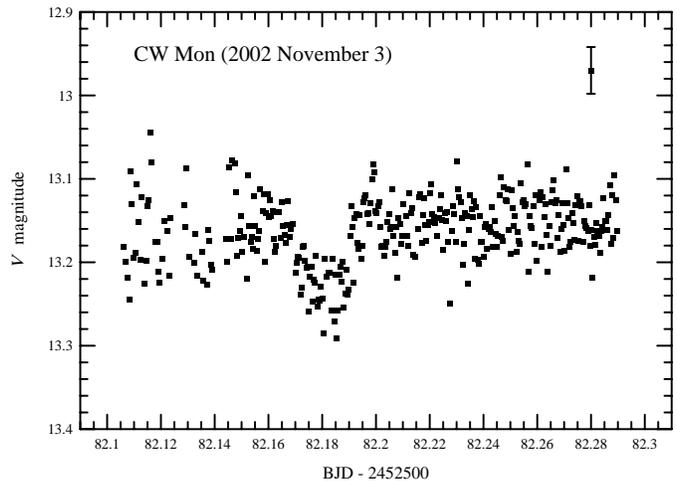}
  \end{center}
  \caption{Eclipse caught on 2002 November 3, when CW Mon was still before
  the outburst maximum.  The shallow depth (0.2 mag) and the eclipse shape
  indicate that the eclipse was a partial one.}
  \label{fig:82}
\end{figure}

\begin{figure}
  \begin{center}
    \FigureFile(88mm,110mm){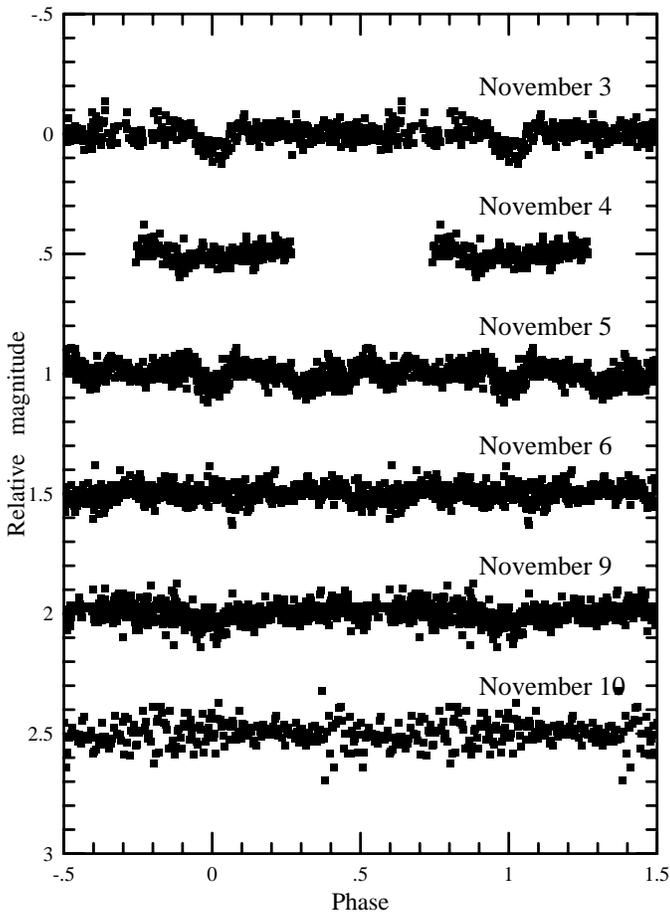}
  \end{center}
  \caption{Nightly folded light curves with a period of 0.1766 d.  The zero
  phase is taken as BJD 2452582.180.  The November 10 data represent the
  averages of 0.005 phase bins in order to reduce the scatter.
  Because of the uncertainty of the adopted period, the phases have
  uncertainties of $\sim$0.07 (November 5) to $\sim$0.22 (November 10).
  }
  \label{fig:fold}
\end{figure}

\begin{figure}
  \begin{center}
    \FigureFile(88mm,60mm){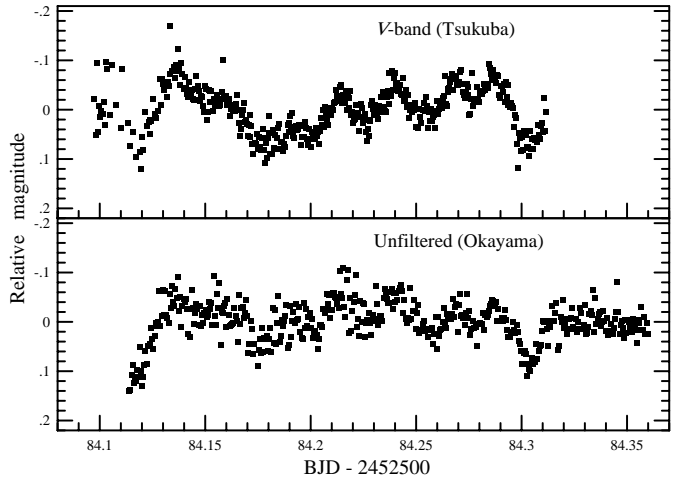}
  \end{center}
  \caption{Short-term variations on 2002 November 6 (around the maximum
  light).  The upper and lower panels represent $V$ and unfiltered
  (close to $R_{\rm c}$) observations, respectively.  Linear trends within
  the night have been subtracted from the observation.
  The fadings around BJD 2452584.117 and 2452584.303 are eclipses
  described in subsection \ref{sec:ecl}.
  The short-term variations are more prominent in the $V$-band, which
  indicates that a hotter region more contributes to the variations.}
  \label{fig:84}
\end{figure}

\begin{table}
\caption{Instruments.}\label{tab:inst}
\begin{center}
\begin{tabular}{cccc}
\hline\hline
Site & Telescope & CCD & Software \\
\hline
Tsukuba (T) & 25-cm SCT & AP-7 & MIRA A/P \\
Okayama (O) & 21-cm refl. & ST-7XE & Java$^*$ \\
Belgium (B) & 35-cm SCT & ST-7 & AIP4WIN \\
Kyoto (K)   & 25-cm SCT & ST-7 & Java$^*$ \\
Hida (H)    & 60-cm refl. & SITe003AB$^\dagger$ & IRAF \\
\hline
 \multicolumn{4}{l}{$^*$ See text.} \\
 \multicolumn{4}{l}{$^\dagger$ PixCellent S/T 00-3194 camera.} \\
\end{tabular}
\end{center}
\end{table}

\begin{table}
\caption{Zero points added to each set of observations.}
\label{tab:offset}
\begin{center}
\begin{tabular}{ccc}
\hline\hline
Date (2002) & Okayama & Belgium \\
\hline
November 5  & 11.85 & $\cdots$ \\
November 6  & 11.78 & 11.74 \\
November 9  & 12.01 & $\cdots$ \\
November 10 & 12.64 & $\cdots$ \\
November 13 & 12.16 & $\cdots$ \\
\hline
\end{tabular}
\end{center}
\end{table}

\begin{table*}
\caption{Journal of CCD photometry.}\label{tab:log}
\begin{center}
\begin{tabular}{ccrcccrccc}
\hline\hline
\multicolumn{3}{c}{Date}& Start--End$^*$ & Filter & Exp(s) & $N$
        & Mean mag$^\dagger$ & Error & Obs$^\ddagger$ \\
\hline
2000 & January  &  7 & 51551.299--51551.314 & none & 30 &  36 & (0.91) &
     0.02 & K \\
     &   &  7 & 51551.401--51551.549 & none & 40 & 285 & (0.76) & 0.01 & B \\
     &   &  8 & 51552.289--51552.328 & none & 30 &  90 & (1.08) & 0.02 & K \\
     &   & 10 & 51554.056--51554.196 & $V$  & 60 & 131 & 12.92 & 0.01 & T \\
     &   & 11 & 51555.034--51555.162 & $V$  & 60 & 101 & 13.01 & 0.01 & T \\
2002 & November &  3 & 52582.106--52582.290 & $V$  & 30 & 351 & 13.16 &
     0.01 & T \\
     &   &  4 & 52583.194--52583.287 & $V$  & 30 & 203 & 13.24 & 0.01 & T \\
     &   &  5 & 52584.097--52584.311 & $V$  & 30 & 433 & 12.82 & 0.01 & T \\
     &   &  5 & 52584.114--52584.360 & none & 45 & 445 & (0.97) & 0.01 & O \\
     &   &  6 & 52584.567--52584.748 & none & 40 & 190 & (1.06) & 0.01 & B \\
     &   &  6 & 52585.090--52585.361 & none & 45 & 492 & (0.98) & 0.01 & O \\
     &   &  6 & 52585.104--52585.285 & $V$  & 30 & 382 & 12.76 & 0.01 & T \\
     &   &  9 & 52588.114--52588.277 & $V$  & 30 & 348 & 13.78 & 0.01 & T \\
     &   &  9 & 52588.117--52588.357 & none & 45 & 367 & (1.68) & 0.01 & O \\
     &   & 10 & 52589.079--52589.366 & none & 45 & 420 & (2.33) & 0.01 & O \\
     &   & 10 & 52589.132--52589.252 & $V$  & 30 & 265 & 14.95 & 0.01 & T \\
     &   & 13 & 52591.776--52591.874 & $V$  & 30 & 216 & 16.07 & 0.03 & T \\
     &   & 13 & 52592.095--52592.142 & none & 90 &  45 & (3.91) & 0.02 & O \\
     &   & 16 & 52595.099--52595.302 & none & 90 & 183 & (4.31) & 0.01 & O \\
     &   & 16 & 52595.212--52595.214 & $V$  & 90 &   3 & 16.40 & 0.03 & H \\
\hline
 \multicolumn{10}{l}{$^*$ BJD$-$2400000.} \\
 \multicolumn{10}{l}{$^\dagger$ Differential magnitudes to the comparison star
                    are given in parentheses except for the Tsukuba and} \\
 \multicolumn{10}{l}{\phantom{$^\dagger$} Hida Observations.} \\
 \multicolumn{10}{l}{$^\ddagger$ T (Tsukuba), O (Okayama), K (Kyoto),
                    B (Belgium), H (Hida)} \\
\end{tabular}
\end{center}
\end{table*}

\section{Observation}

   The CCD observations were carried out at several sites.  The instruments
are given in table \ref{tab:inst}.
We mainly used GSC 146.1617 ($V$ = 11.75, $B-V$ = 0.65, \cite{mis96sequence})
as the primary comparison star, whose constancy during the observation
was confirmed by a comparison with GSC 146.1362 and several fainter
check stars.  The Belgium observation used GSC 146.1677 (Tycho-2\,$V$ = 11.86,
$B-V$ = 1.05) as the primary comparison.  The Okayama and Kyoto images
were dark-subtracted, flat-fielded, and analyzed using
the Java$^{\rm TM}$-based aperture and PSF photometry package developed
by one of the authors (TK).  The Tsukuba, Belgium and Hida images were
analyzed in a similar standard way, with the MIRA A/P, AIP4WIN and
IRAF,\footnote{
  IRAF is distributed by the National Optical Astronomy
  Observatories for Research in Astronomy, Inc. under cooperative
  agreement with the National Science Foundation.
}
respectively.  The Tsukuba and Hida observations were converted to $V$
magnitude scale using the photometric sequence \citep{mis96sequence}.

   Before combining the entire data sets of the 2002 observation, we took
the following procedure.

   Since each observer used a different filter and CCD, we first added
a constant to each set of observations in order to obtain a common
magnitude scale, which was adjusted to the $V$-band Tsukuba observation.
The constants were chosen to maximize the cross-correlation after the
correction (table \ref{tab:offset}).  When there are no overlapping
observations, we calculated the most likely value using interpolation.
Since outbursting dwarf novae are known to have colors close to $B-V=0$,
the difference in the systems would not significantly affect the following
analyses of periodicities and the general outburst properties.
The large difference on November 10 (table \ref{tab:offset}) may represent
a result of an increased contribution of the cool part of the accretion
disk during the outburst decline.  On November 13, this effect became
smaller.

   Barycentric corrections to
the observed times were applied before the following analysis.  The log
of observations is summarized in table \ref{tab:log}.  In the next section,
we first deal with the 2002 observation.  We refer to the 2000 observation
when a comparison is necessary.

\section{Results}

\subsection{Long-Term Light curve}

   Figure \ref{fig:long} shows a long-term visual light curve of CW Mon
constructed from the observations (1996--2002) reported to VSNET.  Large
and small dots represent positive and negative (upper limit) observations,
respectively.  Outbursts rather rarely occur, typically once in 100--200 d.
These observations generally confirmed the results by \citet{stu97cwmon}.

\subsection{Course of the 2002 October--November Outburst}\label{sec:2002out}

   The 2002 October--November outburst was first detected on October
29.378 UT at a visual magnitude of 14.4.  The object was not seen in
outburst on October 28 by three independent observers (VSNET observations).
The object further brightened to a visual magnitude of 12.9 on October
31.315 UT.  These observations indicate that the object was caught during
its earliest rising stage.  We define October 29.378 as $t$ = 0.

   Subsequent CCD observations (table \ref{tab:log}) showed a ``premaximum
halt'' until November 4 ($t$ = 5 d).  After this halt, the object further
brightened to $V$ = 12.7--12.8 on November 6 ($t$ = 7 d).  After reaching
this maximum, the object rather rapidly faded.  The rate of decline
reached 0.7--1.2 mag d$^{-1}$ (between November 9 and 10).
The present outburst is characterized by a rather slow
rise, accompanied by a premaximum halt, and a more rapid decline.
On November 13, the object almost reached the quiescence ($V$ = 16.1),
followed by a further 0.3 mag fade in three days.

   The overall light curve of the 2002 October--November outburst
is shown in figure \ref{fig:out}.

\subsection{Eclipse Detection}\label{sec:ecl}

   On 2002 November 3, we detected a shallow fading (depth 0.2 mag) lasting
for $\sim$30 min (figure \ref{fig:82}).
The light curve was otherwise relatively flat.
Since the properties of the phenomenon closely agree with the description
by \citet{sto87highspeedCCD}, we identified it to be an eclipse
(hereafter we call these phenomena eclipses).
The shallow depth and the eclipse shape indicate that the eclipse was
a partial one, as suggested by \citet{sto87cwmonbaas}.  Assuming a period
of 0.1762 d \citep{szk86cwmonxleoippegafcamIR}, the phenomenon was confirmed
to occur at the same phase on November 5 (cf. figure \ref{fig:fold}),\footnote{
   The quoted error in \citet{szk86cwmonxleoippegafcamIR} corresponds to
   an uncertainty of $\sim$0.1 hr in identifying the phases between
   November 3 and 5.  This uncertainty will not affect the identification
   of the observed phenomena.  A unique identification of the orbital phases
   based on \citet{szk86cwmonxleoippegafcamIR} was impossible because of
   the lack of precision to make a long-term ephemeris.
}
there was no comparable eclipse at the same phase on November 4, 6, 9, and 10.
These observations indicate that the eclipses are essentially transient,
i.e. the eclipses occur only when the accretion disk is large enough to be
eclipsed.  Because of this transient nature of the eclipse phenomenon,
we have not attempted to refine the eclipse ephemeris by
\citet{szk86cwmonxleoippegafcamIR}.

   We determined the mid-eclipse times by minimizing the dispersions of
eclipse light curves folded at the mid-eclipse times.  The error of
eclipse times were estimated using the Lafler--Kinman class of methods,
as applied by \citet{fer89error}.  Since the eclipse profiles and depths
considerably varied, the resultant errors, however, should better be
treated as a statistical measure of the observational errors.  The times
are given in table \ref{tab:eclmin}.  A linear regression of the times
yielded the following ephemeris.  The errors correspond to the epoch
at $E$ = 8.  The resultant period agrees with the value by
\citet{szk86cwmonxleoippegafcamIR} within their respective errors.

\begin{equation}
\rm{BJD_{min}} = 2452582.1801(38) + 0.17659(69) $E$. \label{equ:reg1}
\end{equation}

\begin{table}
\caption{Eclipses and $O-C$'s of CW Mon.}\label{tab:eclmin}
\begin{center}
\begin{tabular}{lrrr}
\hline\hline
Eclipse$^*$ & Error$^\dagger$ & $E$$^\ddagger$ & $O-C$$^\S$ \\
\hline
52582.1801 &  5 &  0 &  4 \\
52584.1174 &  9 & 11 & $-$48 \\
52584.3032 &  5 & 12 &  44 \\
\hline
 \multicolumn{4}{l}{$^*$ Eclipse center.  BJD$-$2400000.} \\
 \multicolumn{4}{l}{$^\dagger$ Estimated error in 10$^{-4}$ d.} \\
 \multicolumn{4}{l}{$^\ddagger$ Cycle count.} \\
 \multicolumn{4}{l}{$^\S$ Against equation (\ref{equ:reg1}).
                           Unit in 10$^{-4}$ d.} \\
\end{tabular}
\end{center}
\end{table}

\begin{figure*}
  \begin{center}
    \FigureFile(180mm,110mm){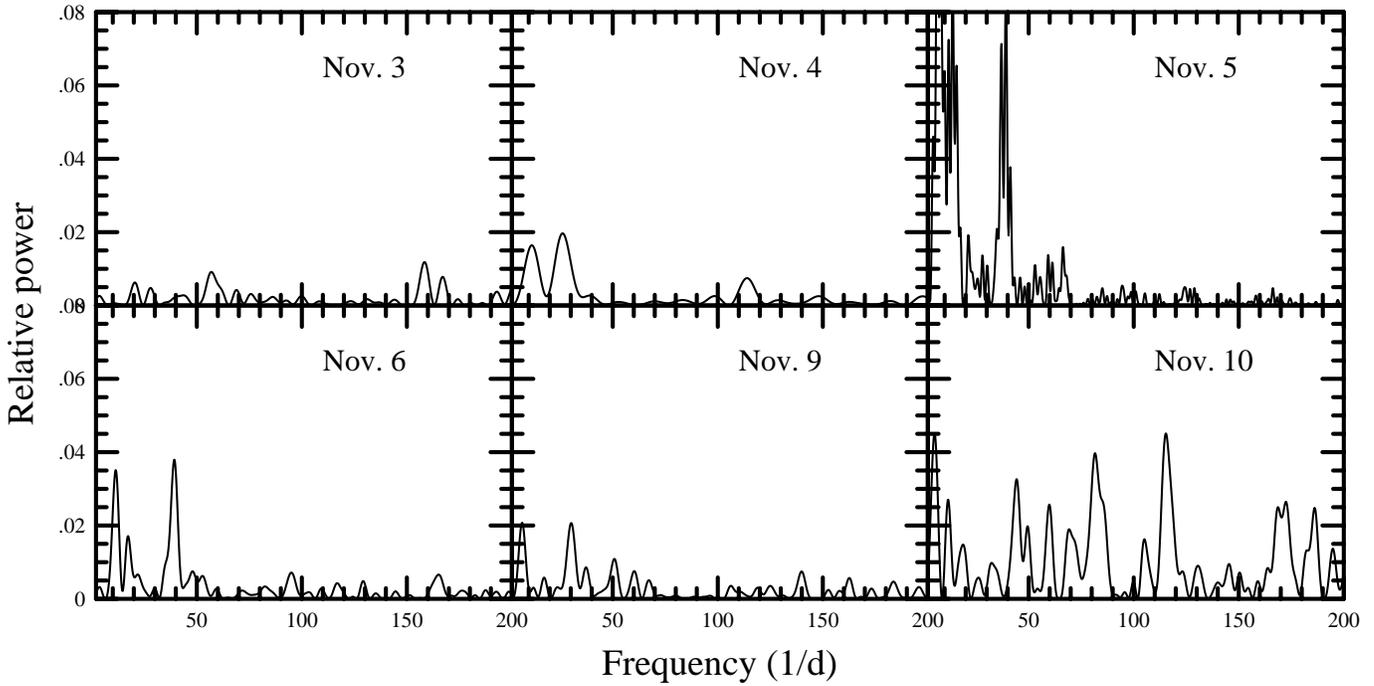}
  \end{center}
  \caption{Nightly power spectra of the short-term variations.  The data
  points around the expected eclipses ($|phase| <$ 0.1 against equation
  \ref{equ:reg1}) were removed before the analyses.  There is a distinct
  power around a frequency of 40 d$^{-1}$ on November 5 and 6.  The power
  of short-term variations was weak on the other nights.  The power spectrum
  on November 10 more reflect a scatter in the light curve rather than true
  signals (cf. figure \ref{fig:pulse}).
  }
  \label{fig:power}
\end{figure*}

\begin{figure}
  \begin{center}
    \FigureFile(88mm,60mm){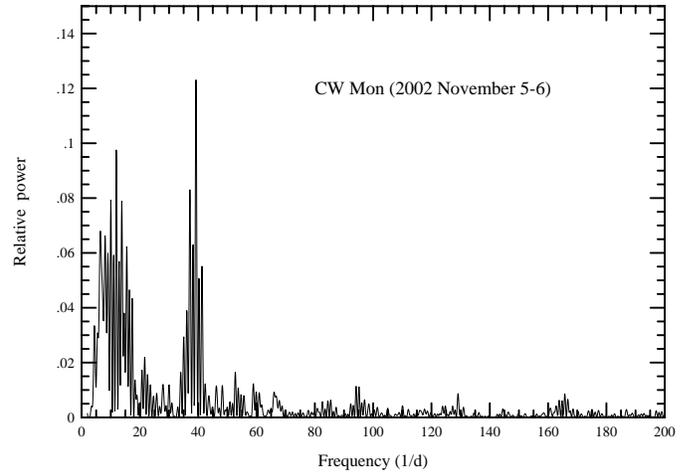}
  \end{center}
  \caption{Power spectrum of the combined data on November 5 and 6.
  A strong coherent signal at a frequency of 39.233(12), corresponding to
  a period of 0.025489(8) d, is clearly seen.
  }
  \label{fig:8485}
\end{figure}

\begin{figure*}
  \begin{center}
    \FigureFile(180mm,110mm){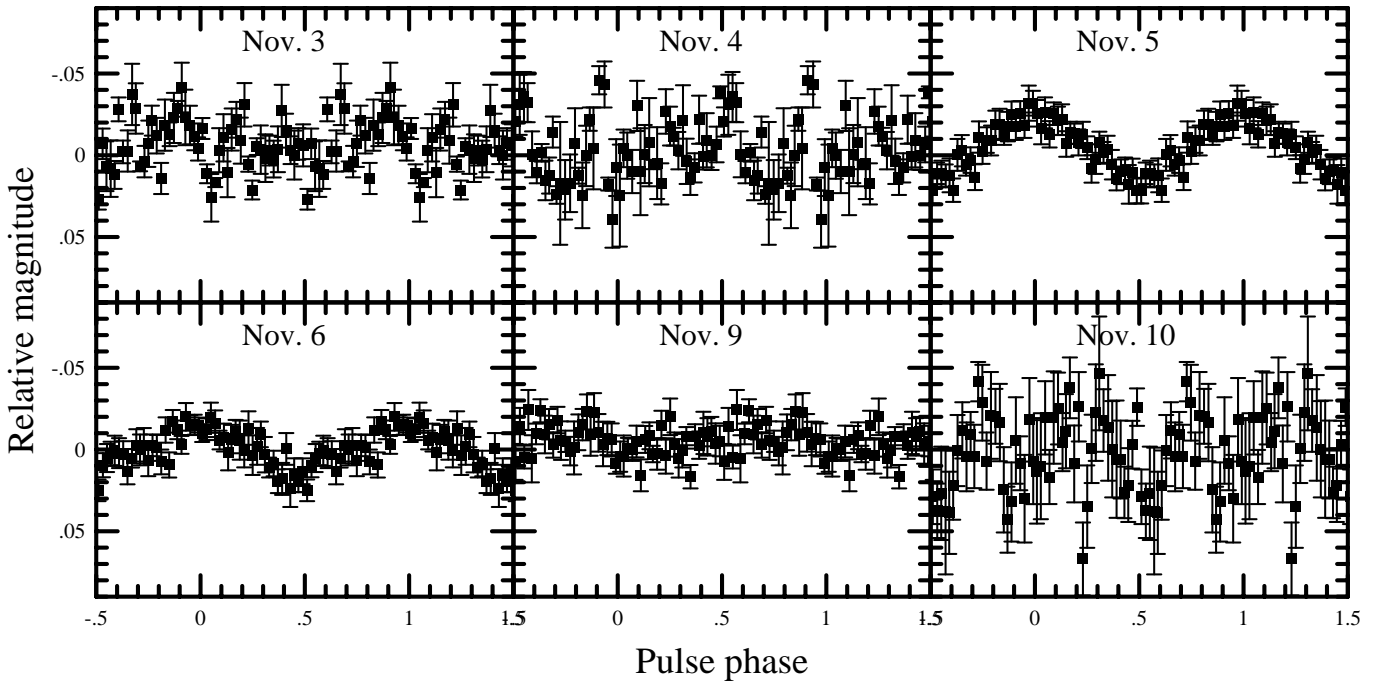}
  \end{center}
  \caption{Nightly pulse profile of the signal at the frequency 39.233
  d$^{-1}$.  The profile is almost sinusoidal on November 5 and 6.
  }
  \label{fig:pulse}
\end{figure*}

\begin{figure}
  \begin{center}
    \FigureFile(88mm,110mm){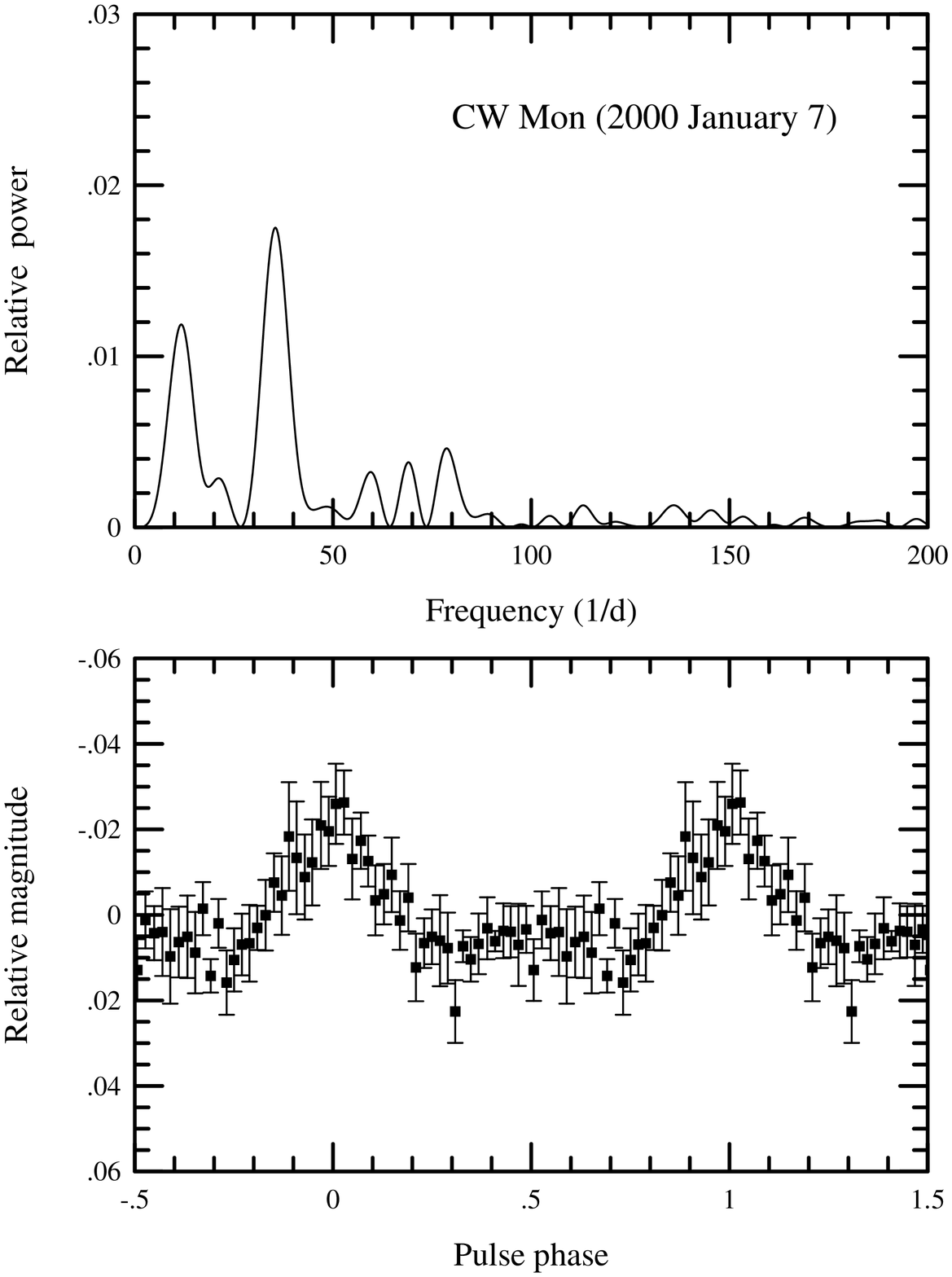}
  \end{center}
  \caption{Power spectrum (upper) and pulse profile (lower) of the 2000
  January 7 data.  The data were taken near the outburst maximum.
  }
  \label{fig:cw20}
\end{figure}

\subsection{Short-Term Variations}

   During the premaximum halt (2002 November 3), the light curve was
rather flat except for the shallow eclipse.  Strong short-term variations
appeared when the object further brightened.  The short-term variations
have typical time-scales of $\sim$40 min.  The amplitude of the short-term
variation reached a maximum around the outburst maximum, and then decayed
as the system faded (2002 November 9 and after).

   Figure \ref{fig:84} shows a comparison of $V$ and unfiltered simultaneous
observations of the short-term variations on 2002 November 6 (around the
maximum light).  The short-term variations are more prominent in the
$V$-band, which indicates that a hotter region more contributes to the
variations.

   Figure \ref{fig:power} shows nightly power spectra of the short-term
variations.  The data points around the expected eclipses ($|phase| <$ 0.1
against equation \ref{equ:reg1}) were removed before the analyses.
Linear trends within each night were also removed by fitting a line.
There is a distinct power around a frequency of 40 d$^{-1}$ ($\sim$ 40 min)
on November 5 and 6 (around the outburst maximum).  The power of short-term
variations was weak on the other nights.

   Figure \ref{fig:8485} shows a power spectrum of the combined data on
November 5 and 6.  A strong coherent signal at a frequency of 39.233(12),
corresponding to a period of 0.025489(8) d, is clearly seen.
Figure \ref{fig:pulse} shows nightly pulse profile of the signal at
the frequency 39.233 d$^{-1}$.  The profile was almost sinusoidal on
November 5 and 6.  The signal was not apparent on the other nights.

   Although there was no premaximum observation during the 2000 January
outburst, the existence of similar short-term variations was confirmed
around or shortly after the maximum light.  Figure \ref{fig:cw20} shows
a power spectrum (upper) and a pulse profile (lower) from the 2000
January 7 data.\footnote{
  Since we do not have an applicable eclipse ephemeris, no rejection
  of the data was applied based on the phase.
}
The data were taken near the outburst maximum.  Although
clear distinction of the period is difficult because of the shortness
of the run, the dominant frequency in the 2002 data was most likely
present.\footnote{
  Preliminary observations after 2000 January 7 (vsnet-alert 3977,
  L. Cook $\langle$http://www.kusastro.kyoto-u.ac.jp/vsnet/Mail/\\
  alert3000/msg00977.htnml$\rangle$ and
  vsnet-alert 3976, S. Walker
  $\langle$http://www.kusastro.kyoto-u.ac.jp/vsnet/Mail/alert3000/\\
  msg00976.htnml$\rangle$) suggest that the short-term variations became
  less coherent on later nights.  This tendency is not inconsistent with
  the behavior seen after the object reached the 2002 November maximum.
}

\section{Discussion}

\subsection{Outburst Properties}\label{sec:outprop}

\begin{figure}
  \begin{center}
    \FigureFile(88mm,110mm){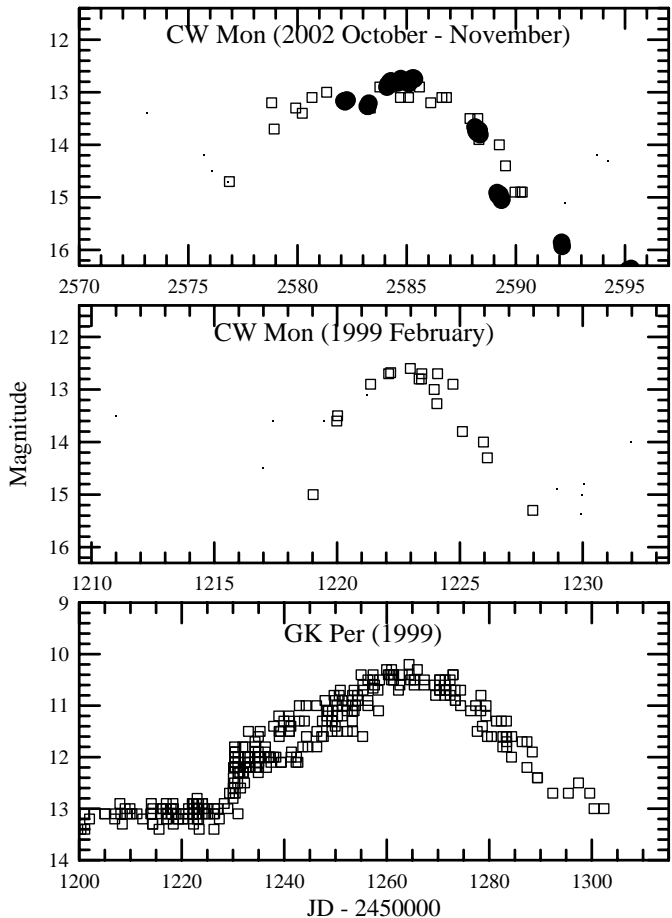}
  \end{center}
  \caption{Comparison of the outbursts of CW Mon and GK Per (the data are
  from VSNET).  The symbols are the same as in figure \ref{fig:out}.
  Note that the scales are different between CW Mon and GK Per.
  }
  \label{fig:outcomp}
\end{figure}

   The outburst properties of CW Mon (subsection \ref{sec:2002out})
are rather unique among dwarf novae with similar orbital parameters.
As a comparison, U Gem ($P_{\rm orb}$ = 0.176906 d, $i$ = 69.7$^{\circ}$
(\cite{krz65ugem}; \cite{war71ugem}; \cite{zha87ugem}), mean cycle
length = 118 d), rarely shows slowly rising outbursts
(\cite{sma93ugem}; \cite{sio97ugemHST}; \cite{can02ugem1985}).
The premaximum halt more resembles those in the outbursts of long-period
dwarf novae with long recurrence times:
BV Cen ($P_{\rm orb}$ = 0.610108 d: \cite{vog80bvcen}; \cite{gil82bvcen};
\cite{men86bvcen},
V1017 Sgr ($P_{\rm orb}$ = 5.714 d: \cite{sek92v1017sgr}),
GK Per ($P_{\rm orb}$ = 1.996803 d: \cite{bia81gkper}; \cite{bia82gkper};
\cite{bia86gkper}; \cite{can86gkper}; \cite{cra86gkperorbit};
\cite{kim92gkper}; \cite{sim02gkper}; \cite{nog02gkper}),
V630 Cas ($P_{\rm orb}$ = 2.564 d: \cite{whi73v630cas}; \cite{hon93v630cas};
\cite{war94v630cas}; \cite{oro01v630cas}).  Figure \ref{fig:outcomp}
presents a representative comparison of the outbursts of CW Mon and GK Per.
The slow rise to the maximum and premaximum halts are common in
these systems.

   Some of infrequently outbursting dwarf novae, such as
CH UMa ($P_{\rm orb}$ = 0.343 d, \cite{bec82chumaXray}; \cite{sim00chuma})
and DX And ($P_{\rm orb}$ = 0.440502 d, \cite{bru97dxand}; \cite{sim00dxand})
sometimes show similar outburst profiles.

   As shown in \citet{kim92gkper}, these outbursts are classified as
a variety of so-called `type B' outbursts (\cite{sma84DNoutburst};
\cite{sma87outbursttype}).  The existence of the premaximum halt
can be well explained as a result of a combination of {\it stagnation}
due to the increase of the specific heat associated with H and He
ionization (\cite{min88uvdelay}; \cite{min90ADirradiation}) and slow
inside-out propagation of the thermal instability starting at the inner
region of the accretion disk.  \citet{kim92gkper} indeed showed a
prolonged stagnation can be achieved in the condition of GK Per.
Since low mass-transfer enables a sufficient quiescent diffusion to
allow inside-out outbursts, and a larger disk radius effectively suppresses
thermal instability to occur in the outer disk, such an condition of
prolonged stagnation can be most easily achieved in long-$P_{\rm orb}$
and low mass-transfer systems, which agrees with the above observational
examples.

   In many short-period dwarf novae with rather high mass-transfer rates,
such a condition is difficult to achieve.
Nevertheless, the same system is shown to undergo unusual outbursts,
resembling those of GK Per, when the mass-transfer rate is temporarily
reduced (RX And: \cite{kat02rxand}).  Many of the well-observed long
outbursts of CW Mon (cf. figure \ref{fig:outcomp}) to some degree bear
resemblance to the 2002 October--November outburst, especially in the
initial rapid rise and the following phase of a slow rise to maximum,
indicating that the stagnation-type progress of the outburst is very
effective in CW Mon.

\begin{figure}
  \begin{center}
    \FigureFile(88mm,60mm){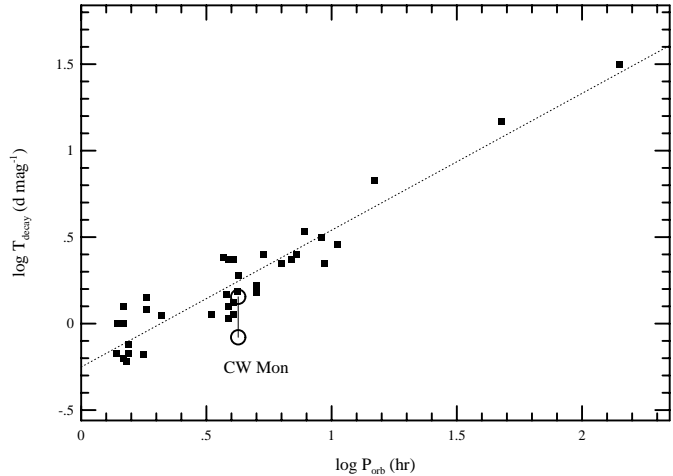}
  \end{center}
  \caption{Relation between the orbital periods and the rates of decline of
  well-observed dwarf novae.  The data and the linear fit are taken from
  \citet{kat02gycnc}.
  The location of CW Mon is shown by the two open circles (upper:
  unfiltered CCD, and lower: $V$), measureed between November 9 and 10.
  The $V$-band decline rate is larger than those of other dwarf novae with
  similar orbital periods.
  }
  \label{fig:bailey}
\end{figure}

   The early decline rates from the outburst maxima (0.3--0.7 mag d$^{-1}$)
do not seem to be inconsistent with the classical Bailey relation
(\cite{BaileyRelation}; \cite{szk84AAVSO}; \cite{kat02gycnc}), indicating
that the early fading part of the outburst in CW Mon follows the standard
time-evolution of dwarf nova outbursts.
At later times, however, the mean decline rate increased to 0.7--1.2
mag d$^{-1}$, which is larger than those of dwarf novae with similar
orbital periods (figure \ref{fig:bailey}).

\subsection{Eclipses}

   As shown in subsection \ref{sec:ecl}, transient eclipses were only
detected on 2002 November 3 (premaximum stage) and November 5
(near maximum).  As is naturally expected from the binary configuration,
these eclipses are interpreted to occur only when the accretion disk
is sufficiently large to be eclipsed by the secondary.  In the model
calculation by \citet{kim92gkper}, the outward heating wave travels toward
the outer disk twice: during the stagnation phase and the maximum phase.
During these epochs, the outer disk can expand as a result of the heating
wave and the resultant upward transition to a hot state
(\cite{ich92diskradius}; \cite{osa96review}).  The detection
of the eclipses only during the premaximum halt (stagnation phase) and
near the maximum light (maximum disk expansion) well agrees with the
picture.  We thus conclude that the present transient eclipse detections
on two occasions are an additional observational support for the
two-step ignition of the outburst, as proposed by \citet{kim92gkper}.

\subsection{Wide and Narrow Outbursts}

   The reported magnitude difference of the wide and narrow outbursts
\citep{stu97cwmon} also seems to support this stagnation-type
interpretation, since no such a great, systematic difference of peak
magnitudes between wide and narrow outbursts has been observed in usual
SS Cyg-type dwarf novae, e.g. SS Cyg
(\cite{can92sscyg}; \cite{can98sscyg}).  Such a difference can be
reasonably reproduced if narrow outbursts represent outbursts with early
quenching of the outward heating wave (for an example of such a simulation,
see \cite{kim92gkper}).

   Among well-observed dwarf novae with similar orbital periods,
TW Vir ($P_{\rm orb}$ = 0.18267 d) shows a similar pattern of
outbursts.\footnote{
  See e.g. $\langle$http://www.kusastro.kyoto-u.ac.jp/vsnet/gcvs/\\
  VIRTW.html$\rangle$
}
In long period systems, such as CH UMa \citep{sim00chuma}, RU Peg
($P_{\rm orb}$ = 0.3746 d), this phenomenon is commonly seen, which is
a natural consequence of the predominant inside-out type outbursts in
these long-period systems.

\subsection{Short-Term Variations}\label{sec:spin}

   From the apparent presence of a short-period coherent signal during
the maximum phase of the 2002 November outburst, and the likely existence
of the common period during the 2000 January outburst, CW Mon is likely an
intermediate polar (IP; sometimes called a DQ Her star: for recent reviews,
see e.g. \cite{kin90IP}; \cite{pat94ipreview}; \cite{hel96IPreview};
\cite{buc00IPpower}; chapters 8 and 9 in \cite{hel01book}).
If this IP-nature is confirmed,
CW Mon is a rare dwarf nova showing IP-type properties during their
outbursts.  The other IPs showing rather regular dwarf nova-like (or
possibly dwarf nova-type) outbursts include DO Dra (\cite{wen83dodra};
\cite{szk02dodra}), HT Cam (\cite{ish02htcam}; \cite{kem02htcam})
and GK Per.  Other IPs showing outbursts include EX Hya \citep{hel89exhya}
and TV Col (\cite{szk84tvcolflare}; \cite{hel93tvcolperiods}).
The outbursts of these objects are, however, more irregular than in
DO Dra, HT Cam and GK Per.

   Up to now, HT Cam is the best-known example which showed a dramatic
increase of the IP pulse strength during the outburst phase
\citep{ish02htcam}.  The pulses of this object are generally weaker in
quiescence (\cite{tov98htcam}; \cite{kem02htcam}).  Following this
analogy, we tentatively identify the 0.025489(8) d as the spin period
of the magnetized white dwarf.  However, we must note that the observed
period is not necessarily the spin period but may be some sort of
quasi-periodic oscillations (QPOs)
(e.g. GK Per, see \cite{mor99gkperQPO}; \cite{nog02gkper}) or beat periods
between the rotation of the white dwarf and the structure or wave in the
accretion disk \citep{war02DNO}.

   Following this interpretation, the dramatic increase of the pulse
strength following the premaximum phase can be best understood as a
result of the dramatic increase of the accretion rate in the inner
disk following the stagnation stage, as is reasonably predicted by
\citet{kim92gkper}.

\begin{table*}
\caption{Astrometry of CW Mon.}\label{tab:astrometry}
\begin{center}
\begin{tabular}{cccccl}
\hline\hline
Source    & R. A. & Decl.               & Epoch & Magnitude \\
          & \multicolumn{2}{c}{(J2000.0)} & & \\
\hline
AC 2000.2 & 06 36 54.470 & +00 02 17.70 & 1909.083 & $b$ = 12.58 \\
USNO A2.0 & 06 36 54.547 & +00 02 17.69 & 1955.881 & $b$ = 16.4, $r$ = 16.0 \\
GSC 2.2.1 & 06 36 54.572 & +00 02 17.19 & 1990.963 & $b$ = 17.50, $r$ = 15.94 \\
2MASS     & 06 36 54.579 & +00 02 17.39 & 1998.723 & $\cdots$ \\
\hline
\end{tabular}
\end{center}
\end{table*}

\subsection{Proper Motion and Distance}

   We examined the available astrometric catalogs (table
\ref{tab:astrometry}).  From a comparison of the positions in
USNO A2.0 and 2MASS catalogs, we detected a probable proper motion of
\timeform{0.57''} $\pm$ \timeform{0.17''} in 42.8 yr.  The position of
GSC 2.2.1 at an intermediate epoch supports its direction.
The Astrographic Catalog (AC) happened to contain this object in
outburst, whose position in the latest reduced version also supports
the proper motion.  The derived proper motion of 13 $\pm$ 4 mas yr$^{-1}$
is consistent with the other estimates of 21.4 mas yr$^{-1}$ (FONAC catalog)
and 13.9 mas yr$^{-1}$ (USNO B1.0 catalog). 

   At a distance of 290 pc \citep{ver97ROSAT}, our proper motion
corresponds to a reasonable transverse velocity of 19 $\pm$ 6 km
s$^{-1}$.  The present proper motion study supports the 290 pc
distance estimate. 
At this distance, the maximum absolute magnitude becomes $M_V$ = +5.4
(assuming the maximum apparent magnitude of $V$ = 12.7, present observation).
This value is significantly fainter
than the mean maximum absolute magnitude of $M_V$ = +4.4, of dwarf novae
with similar orbital periods \citep{sma99DNviscosity}.  The value is
fainter than $M_V$ = +4.5 expected from
Warner's relation \citep{war87CVabsmag}.  Although the high inclination may
be partly responsible for this faintness, the formulation by
\citet{war86NLabsmag} indicates that this effect is less than 0.1 mag
compared to an average inclination ($i$ = 44$^{\circ}$), or 0.4 mag
compared to a pole-on system.  The effect of inclination thus seems to be
insufficient to explain the low maximum brightness of CW Mon.

   As described in subsection \ref{sec:outprop}, (at least some of)
the outbursts of CW Mon are well represented by a combination of
stagnation and inside-out type evolution.  In such a situation,
either heating wave may not fully reach the outermost disk, or the
entire disk may not reach the critical surface density
($\Sigma_{\rm max}$), which is a necessary condition that the maximum
$M_V$ follows Warner's relation \citep{can98DNabsmag}.  The faintness
of the outbursts of CW Mon can be thus naturally explained within the
framework of the disk instability theory.

\section{CW Mon as a Possible Intermediate Polar}

   As shown in subsection \ref{sec:spin}, CW Mon is a good IP candidate
which shows transient IP pulses during the maximum phase of the outbursts.
The supposed IP-nature would naturally explain the rather unusual
properties in this system.  The truncation of the inner disk by the magnetic
field lines of the white dwarf can effectively suppress
disk instabilities in the inner accretion disk, thereby lengthening
outburst intervals (\cite{ang89DNoutburstmagnetic}; see also
\cite{kat02gzcncnsv10934}).  The low maximum luminosity can be alternatively
explained by the inner truncation of the accretion disk.  A slight deviation
from the Bailey relation during the late decline phase (subsection
\ref{sec:outprop}) may be a result of the inner truncation
(cf. \cite{ish02htcam}; \cite{kat02gzcncnsv10934}).

   One of the most remarkable feature of CW Mon would be the relatively
strong X-ray emission.  According to \citet{ver97ROSAT}, the X-ray luminosity
of CW Mon is estimated to be $L_X$ = 10$^{31.1}$ erg s$^{-1}$, which
makes CW Mon one of the luminous sources among dwarf novae.  This value is,
however, a rather common X-ray luminosity among IPs \citep{ver97ROSAT}.
The hard spectrum of the X-ray emission (\cite{ric96ROSATCV};
\cite{ver97ROSAT}) is also consistent with the IP picture.
The observed X-ray luminosity also fits the general relation between
$P_{\rm orb}$ and $L_X$ \citep{pat94ipreview}.

   The ratio between the suggested spin period ($P_{\rm spin}$ = 0.0255 d)
and the orbital period ($P_{\rm orb}$ = 0.1766 d) is close to the expected
ratio ($P_{\rm spin}/P_{\rm orb} \sim$0.1) for the equilibrium spin rate
of the white dwarfs in IPs (\cite{kin93IPblob}; \cite{wyn95IPaccretion}).
The observed ratio, however, is slightly above the equilibrium value;
such a situation is relatively rare in IPs (cf.
\cite{wu91MCVmagneticmoment}; \cite{pat94ipreview}; \cite{hel96IPreview}).
V1025 Cen (\cite{buc98v1025cen}; \cite{hel98v1025cen}) and EX Hya
(\cite{jab85exhya}; \cite{hei87exhyaEinstein}; \cite{bon88exhyaspinup})
are the best-known such examples.\footnote{
  The ratio ($P_{\rm spin}/P_{\rm orb}$) would be closer to 0.1 if the
  observed periodicity corresponds to a beat period between the rotation of
  the white dwarf and the orbital motion or a slowly rotating portion of
  the accretion disk.  We note, however, there is no established IP showing
  a beat period as the predominant signal during its outbursts.
}
\citet{kin99exhyaspin} proposed that
the deviation of EX Hya from the usual equilibrium spin rate can be
explained considering an equilibrium condition when the corotation radius
is comparable with the distance from the white dwarf and the L$_1$ point.
\citet{kin99exhyaspin} suggested that such a condition should usually
require an orbital period below the period gap ($P_{\rm orb} \leq$ 2 hr).
Since CW Mon apparently does not fit this picture, further confirmation of
the suggested spin period is strongly recommended.

   The weakness of the IP signature except during the outburst maxima
may be a result of a lower magnetic field than in other IPs.  A deep search
for ultraviolet and X-ray pulses in quiescence (cf. \cite{pat92dodra};
\cite{pat93dodraXray}; \cite{has97dodraHST}; \cite{nor99dodrav709cas};
\cite{szk02dodra}), as well as detailed optical coverage during future
outbursts, is strongly recommended.

\vskip 3mm

We are grateful to many VSNET observers who have reported vital observations.
We are particularly grateful to Dan Taylor, Mike Simonsen and Hazel McGee
for promptly notifying the outbursts through VSNET, and to Lew Cook and
Stan Walker for making their preliminary results quickly and publicly
available.
We are grateful to Rod Stubbings for making the relevant RASNZ VSS
publications available to us.
This work is partly supported by a grant-in aid [13640239 (TK),
14740131 (HY)] from the Japanese Ministry of Education, Culture, Sports,
Science and Technology.
Part of this work is supported by a Research Fellowship of the
Japan Society for the Promotion of Science for Young Scientists (MU).
This research has made use of the Digitized Sky Survey producted by STScI, 
the ESO Skycat tool, and the VizieR catalogue access tool.

\end{document}